\documentclass[eqsecnum,superscriptaddress,showpacs]{revtex4}
\usepackage{graphics}
\usepackage{latexsym}
\usepackage{bm}
\begin{document}
\title{A New Relation between Lamb Shift Energies}

\author{Hiroaki Kubo}\email{n-kubo@phys.cst.nihon-u.ac.jp}
\author{Takehisa Fujita}\email{fffujita@phys.cst.nihon-u.ac.jp}
\author{Naohiro Kanda}\email{nkanda@phys.cst.nihon-u.ac.jp}
\author{Hiroshi Kato}\email{hhkato@phys.cst.nihon-u.ac.jp}
\author{Yasunori Munakata}\email{munakata@phys.cst.nihon-u.ac.jp}
\author{Sachiko Oshima}\email{oshima@phys.cst.nihon-u.ac.jp} 
\author{Kazuhiro Tsuda}\email{nobita@phys.cst.nihon-u.ac.jp}
\affiliation{Department of Physics, Faculty of Science and Technology, 
Nihon University, Tokyo, Japan}

\date{\today}%

\begin{abstract}

We derive a new relation between the observed Lamb shift energies of hydrogen 
and muonium atoms. The relation is based on the non-relativistic description of the 
Lamb shift, and the proper treatment of the reduced mass of electron and 
target particles (proton and muon) leads to the new formula which is expressed as  
$\displaystyle{   {\Delta E^{(H)}_{2s_{1/2}}\over \Delta E^{(\mu)}_{2s_{1/2}} }
=\left({1+{m_e\over m_\mu}\over 1+{m_e\over M_p}} \right)^3   }$. 
This relation achieves an excellent agreement with experiment and presents an 
important QED test free from the cutoff momentum $\Lambda$.

\end{abstract}


\maketitle

\noindent

\section{Introduction}
The Lamb shift energy in hydrogen atom is a symbol of the success in the QED 
renormalization scheme, and indeed the renormalization effect of the electron self-energy 
is responsible for the small deviation of the $2s_{1\over 2}$ energy 
from the prediction of the Dirac equation. In the theoretical calculation 
of the Lamb shift energy, there is some ambiguity which arises from the cutoff momentum 
$\Lambda$ since the calculation is only possible for the non-relativistic treatment, 
at least, up to the present stage. In the non-relativistic evaluation, the Lamb shift 
energy has a logarithmic divergence, and people take the cutoff momentum $\Lambda$ as 
electron mass, that is, $\Lambda =m_e$, which has, of course, no physically plausible 
reason.  In addition, people normally consider the Coulomb propagator modification as
$$ {1\over q^2} \Rightarrow {1\over q^2}\left(1-{\alpha\over 15\pi}
{q^2\over m^2_e} \right) \eqno{(1.1)} $$
which is discussed in the textbook of Bjorken and Drell \cite{bd}. However, the Coulomb 
field $A_0$ should not be quantized since it is a time independent field \cite{fujita}. 
Therefore, there should not be any change of the Coulomb propagator and the modified 
propagator discussed in Bjorken and Drell is not a correct treatment. 

Here, we present a careful calculation of the Lamb shift energy in hydrogen 
and muonium atoms with the non-relativistic field equations. 
In this calculation, we find a new relation between Lamb shift energies of hydrogen 
and muonium atoms. The relation does not depend on the cutoff momentum $\Lambda$ 
and achieves an excellent agreement with experimental 
observations of Lamb shifts, even though the experimental uncertainty of the Lamb 
shift energy in muonium is still too large to decide which of the theoretical model 
calculations should be preferred \cite{woo}.  

\section{Quantization of Coulomb Fields}
Before going to the discussion of the Lamb shift energy, we should first clarify 
the quantization of the electromagnetic field $A_\mu$. After taking the Coulomb gauge 
fixing of $\bm{\nabla} \cdot \bm{A} =0 $, 
the field equation $ \partial_\mu F^{\mu \nu} =ej^\nu $ can be written 
for the $A_0$ field as
$$ \bm{\nabla}^2 A^0 =-ej^0_e \eqno{(2.1)}  $$
where $j^0_e$ denotes the current density of electron. 
This is a constraint equation and therefore the $A_0$ field can be solved and 
written in terms of the electron current $j^0_e$ as
$$ A^0(\bm{r}) ={e\over 4\pi}\int {j^0_e(\bm{r}')\over |\bm{r}-\bm{r}'|} 
d^3r'  . \eqno{(2.2)}  $$
This means that there is no way to 
quantize the $A_0$ field, even though, in the literatures \cite{bd}, this field is 
often quantized in the same way as the vector field $\bm{A}$. The important point is 
that fields should be quantized only when they are time dependent. The creation and 
annihilation operators depend, of course, on time since they occur at some fixed 
point of time in the reaction processes.  
In terms of the Hamiltonian, the Coulomb interaction $H_C$ can be written as 
\cite{fujita,nishi}
$$H_C=-{e^2\over 4\pi} \int { j^0_p(\bm{r}') j^0_e(\bm{r}) d^3rd^3r'  
\over{|\bm{r}'-\bm{r}| }}    \eqno{ (2.3)}  $$
where $j^0_p$ denotes the proton current density. This expression is independent of 
the gauge choice and it clearly states that the Coulomb interaction is not influenced 
by the higher order corrections since eq. (2.3) is exact. 

\subsection{Uehling Potential}
Here, we should comment on the Uehling potential which is essentially the same 
as the finite term of the vacuum polarization in the Coulomb potential \cite{ueh}. 
Uehling obtained the induced charge distribution due to the creation of electron 
and positron pairs in the vacuum as
$$ \delta \rho(\bm{r})=-{\alpha\over 15\pi m^2_e} \bm{\nabla}^2 \rho(\bm{r}) 
 \eqno{ (2.4)}  $$
which can generate the Uehling potential. However, it is well established by now 
that the creation of fermion pairs can be possible only when the fermion fields 
can couple to the vector field $\bm{A}$. Therefore, when the fermions interact 
with the Coulomb field $A^0$, there is no physical process which can create 
the fermion pairs. As an intuitive picture, one may say that the static Coulomb 
field cannot make any polarizations in the vacuum since the pair creations 
are physical processes which involve time dependent fluctuations of 
the vector field $\bm{A}$.  

\subsection{Classical Picture of Polarization}
This classical picture of the fermion pair creations (Uehling potential) 
should come from the misunderstanding of the structure of the vacuum state. 
In the medium of solid state physics, the polarization can take place 
when there is an electric field present. In this case, the electric field 
can indeed induce the electric dipole moments in the medium, and this corresponds 
to the change of the charge density. However, this is a physical process 
which can happen in the real space (configuration space). 
On the other hand, the fermion pair creation in the vacuum in field theory is 
completely different in that the negative energy states are all filled 
in momentum space, and the time independent field of $A^0$ which is only a 
function of coordinates cannot induce any changes on the vacuum state. Therefore, 
unless some time dependent field is present in the reaction process, the pair creation 
of fermions cannot take place in physical processes. 

Therefore, in contrast to the common belief, there is, unfortunately, no change 
of the charge distribution in QED vacuum, even at the presence of two charges, and 
this is basically because the Coulomb field is not time dependent. 

\subsection{Higher Order Corrections}
Therefore, there are no higher order corrections of the vertex 
corrections and vacuum polarization effects to the Lamb shift energy.  
As a possible effect on the Lamb shift energy, there may be two loop self-energy 
corrections to the Lamb shift energy \cite{yero}. However, before examining 
the higher order corrections, we may have to understand how to control 
the cutoff $\Lambda$ effects in the non-relativistic treatment of 
the Lamb shift energy. 

\section{Lamb Shift in Hydrogen Atom}
Here, we present a standard scheme of the Lamb shift in hydrogen atom which is 
based on the non-relativistic treatment.

\subsection{Non-relativistic Treatment}
We start from the Hamiltonian for electron in hydrogen atom 
with the electromagnetic interaction
$$ H={\bm{\hat p}^2\over 2m_0}-{e^2\over r} -{e\over m_0}\bm{\hat p}\cdot \bm{\hat A}  
\eqno{(3.1)} $$
where the $ \bm{\hat A}^2$ term is ignored in the Hamiltonian. 
In this calculation, the electromagnetic field $\bm{A}$ 
should be quantized 
$$ \bm{\hat A}(x)=\sum_{\bm{k}} \sum_{\lambda =1}^2{1\over{\sqrt{2V\omega_{\bm{k}}}}} 
\bm{\epsilon}(\bm{k},\lambda) \left[ c_{\bm{k},\lambda} e^{-ikx} +
  c^{\dagger}_{\bm{k},\lambda} e^{ikx} \right]  \eqno{(3.2)}  $$
where $c^{\dagger}_{\bm{k},\lambda}$ and $c_{\bm{k},\lambda}$ denote 
the creation and annihilation operators which satisfy 
the following commutation relations 
$$ [ c_{\bm{k},\lambda}, \  c_{\bm{k}',\lambda'}^\dagger  ] = \delta_{ \bm{k}, \bm{k}' }
\delta_{\lambda, \lambda'} \eqno{(3.3)}  $$
and all other commutation relations vanish. 

\subsection{Second Order Perturbation Energy}
Now, the second order perturbation energy due to the electromagnetic 
interaction for a free electron state can be written as
$$ \delta E=-\sum_{\lambda} \sum_{\bm{k}} \sum_{\bm{p}'} \left({e\over m_0}\right)^2
{1\over{2V\omega_{\bm{k}} }} { |\langle \bm{p}' |\bm{\epsilon}(\bm{k},\lambda) 
\cdot \bm{\hat p} | \bm{p} \rangle |^2 \over{ E_{\bm{p}'}+k-E_{\bm{p}} }} 
\eqno{(3.4)}  $$
where $| \bm{p} \rangle $ and $| \bm{p}' \rangle $ denote the free electron state 
with its momentum. Since the photon energy ($\omega_{\bm{k}}=k$) is much larger than 
the energy difference of the electron states $ (E_{\bm{p}'}-E_{\bm{p}})$, 
one obtains
$$ \delta E=-{1\over 6\pi^2}\Lambda \left({e\over m_0}\right)^2 \bm{p}^2 \eqno{(3.5)} $$
where $\Lambda$ is the cutoff momentum of photon. This divergence is proportional 
to the cutoff $\Lambda$ which is not the logarithmic divergence. However, this is 
essentially due to the non-relativistic treatment, and if one carries out 
the relativistic calculation of quantum field theory, 
then the divergence becomes logarithmic. 

\subsection{Mass Renormalization and New Hamiltonian}\index{mass renormalization}
Defining the effective mass $\delta m$ as
$$ \delta m={1\over 3\pi^2}\Lambda e^2 \eqno{(3.6)}  $$
the free energy of electron can be written as
$$ E_F ={\bm{p}^2\over 2m_0}-{ \bm{p}^2\over{2m_0^2}}\delta m 
\simeq  {\bm{p}^2\over 2(m_0+\delta m)}  \eqno{(3.7)} $$
where one should keep only the term up to order of $e^2$ because of the perturbative 
expansion. 
Now, one defines the renormalized (physical) electron mass $m_e$ by
\index{renormalized mass}
$$ m_e=m_0+\delta m  \eqno{(3.8)} $$
and rewrites the Hamiltonian $H$ in terms of the renormalized electron mass $m$ 
$$ H={\bm{\hat p}^2\over 2m_e}-{e^2\over r}+ { \bm{\hat p}^2\over{2m_e^2}}
\delta m -{e\over m_e}\bm{\hat p}\cdot \bm{\hat A}  . \eqno{(3.9)}  $$
Here, the third term ( $ { \bm{\hat p}^2\over{2m_e^2}}\delta m $ ) corresponds to 
the counter term which cancels out the second order perturbation energy [eq.(3.4)].

\subsection{Lamb Shift Energy in Hydrogen Atom}
Now, we consider hydrogen atom, and its Hamiltonian can be written as 
$$ H_0= {\bm{\hat p}^2\over 2m_r}-{e^2\over r}\eqno{(3.10)}  $$
where $m_r$ denotes the reduced mass of the electron and proton system. 
Using eq.(3.9), one can calculate the first and the second order perturbation energies  
due to the electromagnetic interaction for the $2s_{1/2}$ state in hydrogen atom 
$$ \Delta E_{2s_{1/2}}={1\over 6\pi^2}\Lambda \left({e\over m_e}\right)^2 
\langle 2s_{1/2} |\bm{\hat p}^2 | 2s_{1/2} \rangle  
  - \sum_{\lambda} \sum_{\bm{k}} \sum_{n\ell} \left({e\over m_e}\right)^2
{1\over{2V\omega_{\bm{k}} }} { |\langle n\ell |\bm{\epsilon}(\bm{k},\lambda) 
\cdot \bm{\hat p} |2s_{1/2} \rangle |^2 \over{ E_{n,\ell}+k-E_{2s_{1/2}} }}  
\eqno{(3.11)} $$
where the first term comes from the counter term. 
This Lamb shift energy for the $2s_{1/2}$ state can be rewritten as
$$ \Delta E_{2s_{1/2}}=  {1\over 6\pi^2}  \left({e\over m_e}\right)^2 \sum_{n,\ell}
|\langle n,\ell  |\bm{\hat p}|   2s_{1/2}\rangle |^2  $$
$$ \times \int_0^\Lambda dk
{E_{n,\ell}-E_{2s_{1/2}} \over{ E_{n,\ell}+k-E_{2s_{1/2}} }} . \eqno{(3.12)} $$
After some calculations, we obtain
$$ \Delta E_{2s_{1/2}}=   {2m_r^3 \alpha^5\over 3m_e^2}  
\ln \left( {\Lambda\over <E_{n, \ell}>}\right)  \eqno{(3.13)}  $$
where we have neglected the ($n, \ell$) dependence in the denominator when summing 
up ($n, \ell$),  and $<E_{n, \ell}>$ is defined as some average value of the excitation 
energies with respect to the $2s_{1/2}$ state. 
For the cutoff $\Lambda$, people normally take $\Lambda \simeq m_e $, but there is no 
special reason why one should take the $\Lambda $ as electron mass. 

\subsection{Cutoff $\Lambda $ Dependence}
Here, we should note that there is, of course, no way to get rid of 
the cutoff $\Lambda $ in the Lamb shift energy in the non-relativistic 
treatment. The divergence of the Lamb shift energy is inevitably a linear one 
in the free state, and it becomes the logarithmic divergence. If one can obtain 
some results which are convergent, then this means that he must have made some mistakes.

\section{Physical Meaning of Cutoff $\Lambda$ }
As one sees, the calculated result of the Lamb shift energy depends on 
the cutoff $\Lambda $, which is not satisfactory at all. However, it should be 
noted that there is no way to avoid the presence of the cutoff $\Lambda $ as long 
as we treat the Lamb shift in the non-relativistic field equations. 

The important point is that we should understand the origin of the value of 
the cutoff $\Lambda $. This should, of course, be understood if one treats it 
relativistically.

\subsection{Relativistic Treatment of Lamb Shift }
The correct scenario of the relativistic treatment of Lamb shift must be 
as follows. In the non-relativistic treatment, the mass counter term 
is linear divergent. However, if one treats it relativistically, 
the divergence is logarithmic. This reason of the one rank down of the divergence 
is originated from the fact that the relativistic treatment considers the negative 
energy states which in fact reduce the divergence rank due to the cancellation. 
Now, we consider the renormalization effect in hydrogen atom, and if we calculate 
the Lamb shift energy in the non-relativistic treatment, then it 
has the logarithmic divergence as we saw above, 
and this is the one rank down of the divergence. This is due to the fact that 
the evaluation of the Lamb shift energy is based on the cancellation between 
the counter term and the perturbation energy in hydrogen atom. In the same way, 
if one can calculate the Lamb shift energy relativistically, then one should obtain 
the one rank down of the divergence, and this means that it should be finite. 

Unfortunately, one cannot carry out the relativistic calculation of the Lamb 
shift energy because of the conceptual difficulty related to the negative 
energy states of the bound systems like hydrogen atom. 
However, we may understand the physical meaning of the cutoff $\Lambda $ in the 
non-relativistic treatment. The value of the  $\Lambda $ must be chosen 
such that the inclusion of the negative energy states are properly simulated 
in the non-relativistic calculation of the Lamb shift. This indicates that we should 
take the  $\Lambda $ value so as to reproduce the experimental observation when 
we evaluate eq.(3.13), and this is the physical reason of the $\Lambda $ value 
why we can take a finite value of $\Lambda $. 

\subsection{Lamb Shift in Anti-hydrogen Atom}
The structure of the negative energy state should be examined if one can measure 
the Lamb shift of anti-hydrogen atom. In this case, positron should feel the effect 
of the vacuum state which should be somewhat different from the case in which 
electron may feel in the same situation. The Lamb shift energy of the $2s_{1/2}$ 
state in anti-hydrogen atom can be written as
$$ \Delta E_{2s_{1/2}}=   {2m_r^3 \alpha^5\over 3m_e^2}  
\ln \left( {\bar{\Lambda}\over <E_{n, \ell}>}\right)  \eqno{(4.1)}  $$
where $\bar{\Lambda}$ denotes the effective cutoff value of the anti-hydrogen atom. 
If the observed value of the $\bar{\Lambda}$ differs from the $\Lambda $ value, 
then it should mean that there is some chance of understanding the structure 
of the negative energy state in the interacting field theory model. 

\subsection{Higher Order Corrections}
Up to now, there are many calculations which treat the higher order corrections 
to the Lamb shift energy. However, as we discuss in section 2, the Coulomb 
interaction or Coulomb propagator cannot be influenced by the higher order 
effects. The only possible higher order effects must be due to the self-energy 
of electron at the order of $e^4$ level \cite{yero}. However, the Lamb shift 
energy depends on the cutoff $\Lambda$, and therefore there is no chance that 
we can predict the Lamb shift energy to a high accuracy. In this respect, 
we should give up reproducing the absolute value of the Lamb shift energy 
in hydrogen atom with sufficient accuracy.

\section{Lamb Shift in Muonium}  
There are very interesting measurements of the Lamb shift energy of 
$2s_{1\over 2}$ state in muonium ($\mu^+e^-$ system) \cite{woo,oram,bad,jung}. 
This presents an important QED test, and we will compare the experimental value 
with the predictions. 

\subsection{Calculations with Vacuum Polarizations}
The Lamb shift energy of $2s_{1\over 2}$ state in muonium has 
been calculated \cite{lau,bha,eri,yen}, and the predicted value becomes 
$$ \Delta E^{(\mu)}_{2s_{1/2}}= 1047.5 \ \ \ \ {\rm MHz} $$
where the main deviation from the hydrogen atom case is due to the vacuum 
polarization contributions. 

\subsection{Prediction with New Relation}
On the other hand, we see that the Lamb shift of muonium can be related to 
that of hydrogen atom as
$$ \Delta E^{(\mu)}_{2s_{1/2}}= \left( {m_r^{(\mu)}\over m_r^{(H)} }\right)^3 
 \Delta E^{(H)}_{2s_{1/2}} \eqno{(5.1)}  $$
where $m_r^{(\mu)}$ and  $m_r^{(H)} $ are given as
$$ m_r^{(\mu)}={m_e\over 1+{m_e\over m_\mu} }, \ \ \ 
m_r^{(H)}={m_e\over 1+{m_e\over M_p} }. \eqno{(5.2)}  $$
Here, $m_\mu$ and $M_p$ denote the masses of muon and proton, respectively. 
Using the experimental value of the hydrogen Lamb shift energy
$$ \Delta E_{2s_{1/2}}^{(H)}(exp) = 1057.862 \ \ \pm0.020 \ \  {\rm MHz} \eqno{(5.3)} $$ 
we can predict the Lamb shift energy of muonium
$$ \Delta E_{2s_{1/2}}^{(\mu)}(th) = 1044  \ \  {\rm MHz} . \eqno{(5.4)} $$ 
This value should be compared to the observed value \cite{woo}
$$ \Delta E_{2s_{1/2}}^{(\mu)}(exp) = 1042 \ \ \pm 22  \ \  {\rm MHz}. \eqno{(5.5)}  $$ 
As can be seen, the agreement is remarkable, and the important point of the prediction 
by the new relation is that it does not depend on the cutoff $\Lambda$.  
However, the experimental accuracy 
is not yet sufficient to decide which of the model calculations is preferred. 
In this respect, it should be very important to improve the observed accuracy 
of the Lamb shift energy in muonium since this should be a very good test of 
the QED renormalization scheme.  

\ 

\section{Conclusions}
We have presented a new relation of the Lamb shift energies between the hydrogen 
and muonium systems, which gives a good test of the QED renormalization scheme. 
This is quite important since the relation does not depend on the cutoff $\Lambda$. 
This clearly shows that the renormalization scheme is indeed a correct theoretical 
framework even though the absolute magnitude of the Lamb shift energy cannot be 
properly predicted by the non-relativistic treatment. 

The new relation can predict the  Lamb shift energy of muonium, 
and the observed Lamb shift energy of muonium is consistent with the prediction, though, 
at present, the experimental uncertainty of the measurement in muonium is still 
a bit too large. However, if the error bar of the observed value can be improved, 
then there is a good chance that the muonium Lamb shift energy can give a stringent 
test of QED renormalization scheme.

\end{document}